# Reconciling complex organizations and data management: the *Panopticon paradigm*

Eric Buffenoir[1]  Isabelle Bourdon[2]

[1]UMR5221, CNRS, Montpellier, France   [2]MRM, Université Montpellier 2, France

*Abstract* These last years, main IT companies have build software solutions and change management plans promoting *data quality management* within organizations concerned by the enhancement of their business intelligence system. These offers are closely similar *data governance schemes* based on a common paradigm called *Master Data Management*. These schemes appear generally inappropriate to the context of complex *extended organizations*. On the other hand, the *community-based data governance schemes* have shown their own efficiency to contribute to the reliability of data in digital social networks, as well as their ability to meet user expectations. After a brief analysis of the very specific constraints weighting on extended organization's data governance, and of peculiarities of monitoring and regulatory processes associated to management control and IT within these, we propose a new scheme inspired by Foucaldian analysis on governmentality: the Panopticon data governance paradigm.

## Introduction

Ten years ago, TDWI (The Data Warehousing Institute) estimated at $ 600 billion the cost of erroneous data in business sector. In fact, data quality control within an organization is a key requirement for the implementation of management control and business intelligence. This question is all the more significant in the extended and complex organizations where differentiation between actors and organizational methods, as well as importance of external influences, strongly constrain the methods adopted to ensure consistency of standards and processes. To deal with issues of data governance, there are currently two major paradigms: Master Data Management and Community Management. The first occupies a market estimated by Gartner to $ 1.9 billion in 2012, up 21% compared to 2011 and 3.2 billion in 2015, and it is difficult to overestimate the markets covered by data quality management inherited from digital social networks.

After defining the global characters of extended organizations and clarified the specific issues of their data governance schemes, as well as the nature of the monitoring and control processes encompassed by the deployment of such governance, we address the legitimacy of existing paradigms (*MDM* and *Community*) in this context, and suggest guidelines for the development of a new data governance paradigm, better suited to the specific challenges addressed by extended organizations.

# 1 Challenges of data management within extended organizations

## 1.1 Management within extended organizations and Information Technologies

These last decades, observers have noticed and studied a large reconfiguration of organizations and the development of new organizational forms as *network [1], virtual [2–4]* or *extended [5], [6]* type. These hybrid organizational forms are characterized by both elements of market and hierarchy *[7], [8]* and are complex organizations that rely on large and open networks *[9]*. Complex organizations combine three types of interdependence *[10]*: interdependence with its environment, interdependence with its own components and finally interdependence between its own components. This complexity is a reflection of the complexity of their internal environment, consisting of processes and technologies at the core of the activity of the organization, as well as a response to their influential external environment including clients, customers, markets, funding suppliers, competitors, and the institutions to which organization must respond *[11], [12]*. The notion of *extended organization or company [6], [13]* is characterized by the existence of multiple relationships with external partners, the delicate definition of its organizational boundaries, which become very porous, tremendous complexity of the causal dynamics in their inner evolutions, as well as nested control processes linking their various entities.

*Structural differentiation* within extended organizations creates a peculiar need for extensive integration of their activities, which can be fulfilled by the development of transverse mechanisms and tools, crossing hierarchical chains and control, and development of multiple control channels for any process *[11], [14]*. The importance of *networks* in the development of cross-integration mechanisms has been extensively studied in the literature on organizational communication *[15]*. The social network theory distinguishes three types of networks and relationships, crisscrossing all organizations and all social systems, based respectively on *woven ties (natural networks), controls (functional networks) or transactions (utility networks) [16]*, one type being always dominant in any network over the other two. The relationships woven within networks may be preeminent over organization's hierarchical controls, due to the possible weakness of this hierarchical power on actors involved in these networks, as being exposed to strong external influences or motivated by their own interests.

Previous considerations results in a set of consequences for the extended organizations we wish to clarify:
- The missions of extended organizations spread across its subunits, its funding mechanisms and the service delivery of a large range of actors to its benefit are very diverse;
- The scope of activity of the organization[1] exceeds that of its own hierarchical authority;
- The relationships based on controls, transactions or ties woven within intra- inter- or trans- organizational networks may be prominent on the hierarchical relationships of the organization;

---

[1] persons or structures benefiting from their services, receiving resources from them, depending partly on their administrative responsibility and whose activities generate products and costs which have to be measured by the organization, possibly shared with partner organizations

- The contractual relationships with other organizations may impose a joint management control of certain shared activities or entities;
- Independent institutions may produce evaluations of its activities, based on external data sources, and hanging over its funding and reputation.

Recent developments in Information Technology (IT) and Information Systems (IS) provide new perspectives for dealing with complex organizational transformations *[17]* and organize their intra- and inter-organizational processes in a dynamic strategic alignment *[18–20]*. Information Systems have a major role to play in strategic management of organizations *[19]*, to support their management control system *[21]* and coordinate their various subunits *[22–24]*. Nevertheless, extended organizations require very specific monitoring and control tools to take into account their previously mentioned distinctive characteristics. In fact, the complexity and diversity of these organizations make it difficult and costly to fully adapt existing tools to their requirements and few tools are able to offer a management control and decisional framework for the organization as a whole by integrating all entities *[24]*.

These tools have to process information from different entities in a multidimensional view, the design of the organizations evolving along with information technologies *[25]*. The organization's extensive control relies increasingly on complex intra and inter-organizational IT systems *[22]*. These IS have tremendously evolved since the 90s, firstly through a deployment of re-engineering, followed by a period of deployment and implementation of ERP *[17]* which has registered limited success *[26], [27]*. Thus, many IS tools have been distributed in organizations to support reorganization processes and management projects such as ERP, CRM, PLM, SCM, DSS, BI. The integration of these complex IS, rapidly raised issues of consistency relying on the need to use the same data set for all operational applications...

### 1.2 Issues of data governance within extended organizations

Deployment of IT puts data, their collation, processing and dissemination issues at the heart of operational management control and decision-making activities *[1]*. Taking into account the development needs of data quality, relevant for both financial and non-financial purposes *[28–30]*, has undergone major changes in the behaviour of actors with respect to the data *[31]* and resulted in a deep redefinition of business processes and partnerships *[19], [21], [32]*.

The data underlying management and decisional processes of most organizations are of various types[2] and cover a wide spectrum. D*ata quality* definition has been the subject of intensive research *[35–40].*

Shortly, the issues raised by a collection of data coming from a set of business areas are:
- loss of traceability concerning the content of data models, the meaning or the legitimacy of values assigned to data ;
- divergence between various data models, coming from different business spheres, concerning the same object ;

---

[2] We generally distinguish *[32]* unstructured data (associated to unnormalized documents), transactional data (registered from interactions between actors), *metadata* (fixing norms for other data and for processes), *master-data* caracterizing the different entities of the organization (agent, resources, administrative bodies, clients, partners,...) which can be spread across different domains *[33], [34]*, and *hierarchical data* (fixing relations between other datas inherited from organizational structure)

- differing values held by different applications for the same data ;
- weak quality of data (duplicates, incompleteness, ...) *[40]*.

The processes systematization devoted to ensure data quality requires manager to develop data governance adapted to this purpose *[41], [42]*. Data Governance formalizes the allocation scheme for rights and duties concerning the use and the management of data within organizations *[42–45]* and encompasses everything that can optimally deal with quality, availability, safety and compliance of data with regulations and standards *[45], [46]*.

Data governance scheme offers a framework for the definition, distribution, synchronization and exchange of reference values for Master Data. These data are generally stored in a single place of reference, which remains in access by different applications, and:
- allows their creation or modification by different actors of the organization,
- ensures its consistent use by various operational applications,
- fixes a set of quality standards,
- facilitates the adaptation to changes of usage patterns,
- allows the construction of relationships between heterogeneous Master Data for decision-making processes.

Unlike data mentioned in a limited number of computer applications or business areas, transverse Master Data especially requires a rigorous treatment within data governance scheme, taking into account issues of IS and of the concerned business areas *[47]*.

The implementation of a data governance scheme aims to generate value for different business areas, by improvement data quality and enrichment of the informational spectrum covered by these data. It necessitates *[47]*:

- semantic alignment between domains,
- clarification of concepts and identification of business glossaries,
- precise definition of business processes,
- identification of control authorities, roles and responsibilities.

We believe that the actual nature of extended organizations imposes a set of technical and organizational constraints on the chosen paradigm of data governance and on the considered IS architecture, reflecting a strong incentive for *decentralization* of control processes over Master Data, although this decentralization may take different forms *[48]*.

- The inherent complexity of extended organizations results in a singular complexity and a wide spectrum of Master Data, reflecting the diversity of actors, missions and organization modes for its subunits. Data governance must promote *deconcentration [48]* to respect the jurisdiction of actors, and *multiplication / diversification of control channels* on a same data set.
- Some communities within the extended organization may prefer to use their proper IS. Other inter- or trans-organizational communities may prefer integrate themselves in data governance schemes held by partner organizations and relying on their own IS tools, rather than adopting the tools and integrate the scheme coordinated by the extended organization. Hence, the pattern of data governance held by the extended organization must allow the *decentralization* of a significant part of control processes towards these communities and partner organizations. Considered *decentralization* is conceived in terms of *functional decentralization or delegation [48],* based on the contractual relationship between the organization and its partners, rather than in its most extreme form of *devolution [48].*

- Certain business processes encompassed by the data governance scheme of the extended organization inevitably involve numerous actors favouring the relations they have woven within networks over the hierarchical controls of the extended organization. The limited efficiency of these control processes does not mean the lack of normative communication concerning the quality of data among the users of these data, but rather a lack of formalization of these processes through tools, standards and processes that underpin the organization's data governance framework. This formalization may be based on the development of digital social networks and their integration in the pattern of data governance.
- Importance of external influences on the activities and resources of the extended organization constrains it to adopt standards for its data repositories that are prepared to the confrontation with the information harvested from relevant external data sources. The lack of control by the organization on IS tools used by the external data sources, imposes a systematic implementation of dictionaries between organization's Master Data and data coming from external sources. The *ranking* of sources appears necessary when facing a deliberate choice of using multiple sources of data.

The development of a data governance paradigm, suitable for extended organizations, raises the question of the precise nature of nested control and regulation mechanisms inherent in the use, the share and the management of data.

### 1.3 The panopticism as a data governance paradigm

The study of mechanisms of monitoring and regulation underlying management control systems and information systems has been the subject of an abundant literature. The coexistence of centralized control and empowerment of actors has been analysed in the studies of control mechanisms underlying the implementation of ERP *[49], [50]*. These studies pointed the proximity of these mechanisms and those of the ideal control paradigm represented by the Panopticon architecture devised by Jeremy Bentham *[51]* and developed by Michel Foucault *[52]*. In this diagram, the actor is placed in a permanent and omnipresent *area of visibility*, is fed continuously to act as if he was being surveiled, and integrate the *norms* and *discipline*. The panopticism is a power that does not need to manifest itself physically, to become effective. Too rapidly identified with a regime of generalized coercion system imposed by a central authority, the panopticism is quite different from living "within a disciplinary system" *[52].* The panopticism is "a general formula that characterizes a type of government" *[53]*. It is a "machinery that assures dissymmetry, disequilibrium, difference. Consequently, it does not matter who exercises power. Any individual, taken almost at random, can operate the machine" *[52]*. A second interpretation of the Panopticon is then summarized by Foucault: "The Panopticon is the formula of liberal *governmentality*," "this new governmental rationality is solely concerned by *interests* and aims at manipulating them" *[53]*.

The data governance paradigms within extended organizations raise a double perspective clarified by Michel Foucault: "the norm is something that can be applied to both a body one wishes to discipline and a population one wishes to regularize. The normalization society is not, in these circumstances, a kind of generalized disciplinary society whose disciplinary institutions have spread and eventually covered the entire space. The normalization society is a society where norm of discipline and norm of regulation intersect along an orthogonal articulation" *[54]*. The data governance paradigm within extended organizations is intended to make the considered organization a social and informational space, subject to omnipresent gaze and regulatory mechanisms, data governance has to deploy "a better and better controlled - more and more rational and economic - adjustment between productive activities, communication networks and the interplay of power relations", it has to develop "a program of governmental rationality… to create a system of regulation of the general conduct of individuals whereby everything would be controlled to the point of self-sustenance, without the need for intervention" *[55]*. This governance lies in "structuring the *field of action* of any individual by every possible ways to influence *representations*, which will play a role in the calculation of their *interests*", by acting on "*monitoring interfaces*". In this way, panoptic power maximizes its action that is to "conduct the conducts" *[53], [55]*.

The use of foucaldian analysis for the study of data governance paradigms imposes a shift of the standpoint concerning Information Systems and some completions of Foucault's analysis to consider peculiarities of Information Technologies.

Michel Foucault focuses his studies on *institutions* in their specific ability to fix individuals in "a place and a collective body there is no way to leave" *[56]*. The control processes promoted through Information Systems can be studied using this peculiar perspective on institutions as soon as we clarify some peculiarities. The nature of information technology is to associate to objects or individual their *digital dual* or avatar registered in databases to proceed prescribed analysis and data matching between heterogeneous data *[57–59]*. The construction of basic business processes within the organization depends so critically on the form chosen for these digital *representations*, that the decision to develop control processes, as well as fields and methods of this control, prove to be consequences of the choice of standards and IS tools within the organization *[60]*. The digital dual is obediently and indefinitely usable for simulations coordinated by the control schemes *[61]*, as real individual is fixed to stay within foucaldian institutions. This *dividualization* takes then place with the consent of the real actors, driven by their interest in the use of digital tools and in the benefits of this simulation *[58]*. The participation of an actor to the control processes devoted to qualify data, relative to him and his environment, is motivated by its need to constitute himself as a subject, which takes shape through an act of recognition of its digital dual. This act of recognition is proceeded each time the actor is "*interpellated*" by the system (in the sense of Althusser's "interpellation" *[62]*) through monitoring interfaces provided by user's personal numeric environment. The precise form of these interfaces impacts deeply the efficiency of the system *[63]*. Previous analysis follows the same singular methodological approach Michel Foucault adopts, by refusing to consider institutions as being primitive objects, fixed prior to any considerations at the same time than the collective body of individuals and their governing rules. Institutions are considered as focal points for the concentration of these control technologies and the production of norms, which are immediately generalized to the whole social body and circulate through a network woven between them, the *subject* resulting from a multiplicity of *subjugation arrangements* within them.

The conceptual framework offered by Foucault appeared very fruitful to analyse the peculiar role played by visibility, transparency and accountability of actors in the deployment of new forms of control mechanisms permitted by IT within organizations

*[64–66]*. It is tempting to reduce Information Technologies to a global realization of the Panopticon control technology, considering the working and living environment of each individual as a space of absolute visibility for their activities *[67], [68]*, and making of IT powerful tools to promote a disciplinary power over individuals, along the lines of Foucault's early works on Panopticon *[52]*. However, this interpretation of digital environments in terms of disciplinary power should be clarified in its singularity. Firstly, prison discipline and disciplines within traditional organizations are immeasurable, one being of a moral nature while the other is of an instrumental one *[69]*. Then, the isolation of the individual at the heart of the Panopticon, which makes of him "the object of information, never the subject of communication" *[52]*, is not that of the individual placed within *area of visibility* created by organization's Information System. The development of *social networks* makes him an actor of transverse communications, eventually diverting information, originally devoted to institutional control, for the purpose of strengthening the resistance of individuals to central authority *[70]*. Starting in the late 80s, it was recognized how the work on Information Systems and management control ignored issues of power and conflict within organizations, and treated organizations as unified entities whose objectives are well defined and widely accepted *[71]*. Resistance to the deployment in extended organizations of control processes underlying ERP has recently been analysed along the lines of Michel Foucault's writings on power and resistance within institutions *[72–74]*. Foucault's studies lead naturally to refuse the standard framework to analyse how the norms and the data governance scheme promoted within an organization can be legitimated and reinforced, or totally changed for another ones. They stand a critical method to analyse the *transformation* of control processes, which disregards schemes/institutions and their rational discourse on their own and privileges the study of elementary disciplinary mechanisms within them and their articulation/discrepancy with the discursive practices *[75],* it finally suggests to clearly distinguish : the rationality/purpose of the governance scheme, the eventually unanticipated effects of it, the positive usage of these effects, and the formalization of a new globalizing rationality/purpose made possible by this usage and absorbing it *[76]*. Our work will analyse the existing data governance paradigms and propose guidelines for a new paradigm directly inspired by previous considerations.

## 2   Towards a data governance scheme adapted to extended organizations

The preceding analysis has led us to present the issues of data governance in extended organizations, in the light of *Panopticon paradigm*. We propose to analyse the specific characters and shortcomings of the existing paradigms of data governance, *MDM* and *Community paradigms*, in the light of our analysis of the challenges addressed by the complexity of these organizations. We thus provide a preliminary analysis of a new data governance paradigm adapted to extended organizations, rooted in our conceptual analysis of the mechanisms of control and regulation within them.

### 2.1  Nature and shortcomings of the MDM Data Governance paradigm

The IT market devoted to data quality has grown through a series of relatively similar strategies and offers, entering the category of schemes called *Master Data Management* *[33], [34]*. Some authors see this one-sided logic as reflecting a new "fashion" for data integration, which follows earlier initiatives on ETL technologies and Data Warehouses, tinged with an ERP-flavoured rhetoric *[77]*.

The MDM includes all operations required by creation, modification or deletion of Master Data *[33]*, and in particular modelling, distribution, quality management, maintenance and archiving of Master Data *[33], [34], [78]*. The main challenge of MDM paradigm is to develop and/or strengthen processes of quality management (cleaning, de-duplication, ...) as systematically as possible *[78]*. Thus, the analysis of business processes of the organization is a prerequisite for the implementation of this scheme *[34], [45], [79]* because the control channels, activated by a proposition to modify a Master Data, rely on the identification of *data-stewards* *[34]* with the required jurisdiction and level of responsibility to provide a level of truth to this proposal and to authorize ultimately its writing as a Master Datum *(golden record)*.

The deployment of MDM systems requires the complete support of the top level managers of the organization, which should drive the whole organization within the logic of improving quality of data in order to involve all stakeholders *[45]*. The success of this deployment relies on the very strong assumption that organizations are homogeneous and highly hierarchically structured. Thus, the MDM scheme relies on:

- the identification of a set of stable-over-time business processes ;
- the clear and precise identification of roles and responsibilities of a limited number of *data-stewards, data-owners* and *data-committees*, placed under the hierarchical authority of the organization, adhering to data quality issues, and inheriting the required jurisdiction to ensure the validation process *[34], [42], [78]* ;
- the direct control on IT tools and master databases (rights for READ,WRITE and ADMIN) used by digital services and operational applications, as well as the use of an integrated digital environment, in order to systematize the dissemination of golden records across applications and enable communities to measure the real-time impact of the control processes on data.

The very nature of extended organizations make difficult the reorganization of *Business Process Management* (BPM) and therefore the application of the MDM scheme within them, because of
- the diversity and instability of their business processes ;
- the inefficiency of hierarchical authority over some elements of control channels promoted by the BPM, because of the prominent influence of networks and external environment on many actors involved in these processes ;
- the low adhesion of middle managers to issues of data quality, due to the illegibility of the consequences of this discipline for their activity, resulting in a refusal to train their teams to the issues and practices of BPM *[74]* ;
- the existence of resistance strategies from senior manager to circumvent the discipline of BPM *[74]* ;
- the difficulties posed by the lack of control by the organization on large parts of IS relied on by its activities, and the multiplicity of IT tools and databases increasingly fragmented by new implementations of IS *[80], [81]*[4] ;
- the difficulties posed by the establishment of data exchange protocols with partner organizations on a suitable collection of data, due to the heterogeneity of the missions and management processes of different partners ;
- difficulties posed by the integration of data harvested from external sources.

---

[4] Dahlberg et al. *[80]* analyses the case of a multinational company whose IS is based on 54 ERPs and reference databases on the 5 continents

While the *MDM paradigm* has nowadays established a monopolistic position on the market of data quality *[45], [82]*, it suffers from its inability to deal with complexity inherent to extended organizations. Moreover, the very specific problems encountered by this model begin to be analysed *[80], [81], [83]*. To our point of view, another approach is needed in the way control processes are promoted by data governance scheme in extended organizations.

## 2.2 Nature and shortcomings of the Community Data Governance paradigm

Adopting a completely opposite philosophy, another paradigm of data governance has taken a prominent place in recent years: the community paradigm. Internet has indeed favoured the development of new forms of collaboration and interaction between actors, facilitated by the manipulation of artifacts and shared information spaces *[84]*. This developing model of collaborative community relies on self-organized online communities, oriented towards the creation and sharing of knowledge *[85], [86]*.

The main examples of collaborative projects based upon virtual communities are open source software (OSS) *[87], [88]* and the development of Wiki technologies. Wiki Technologies allow collaborative, open, egalitarian and anonymous publishing and editing processes of data *[89]*, using mechanisms that track revision history *[90], [91]*. One of the best-known applications of Wiki systems is the collaborative online encyclopaedia Wikipedia [92–96]. Hansen et al. *[97]* recognize to data governance paradigm underlying Wikipedia the ability to offer the conditions of the *Habermasian* ideal type of *rational discourse* for the communication between actors *[98]*, and thus to promote *group rationalization* and *emancipation of individuals* *[99]*.[5]

The systems whose data governance model relies on this paradigm are recognized to produce data of a remarkable quality in a rather short time [92], [100–103]. This data quality is also a major source of value for these organizations. The final data (or its latest version) is the product of a social interactions process, embodied in the iterative and negotiated changes on a selected collection of data, between actors *[101]* within a virtual community [104–106].[6] However, the last individual to edit a set of data defines the reference value for these data, the reliability criterion for a set of data relies then primarily on the observed lack of conflict over these data within the concerned virtual community. This pattern of data governance differs greatly from *centralized disciplinary systems* based on MDM paradigm; it relies on a *democratic relativism* philosophy *[107]*, indeed:
- it is based on a relativistic principle imposing the neutrality of the point of view ;
- it ensures an egalitarian distribution of rights of exclusion and expression within affected communities ;
- it allows communities to establish their own control channels, allocate autonomously the levels of rights and responsibilities, and evaluate these choices in a collegial process.

The critical facts promoting quality of contents are firstly related to the system's ability to gather broad communities and attract strong "diversity of cognitive experiences" and of knowledge of their members *[96]*, along the lines of the analysis on *crowds [108].* As shown by the literature on *Group Decision Support Systems*, anonymity is an undeniable incentive for the engagement of numerous stakeholders in a critical practice *[109]*, the quality of contents being guaranteed by the existence of regulatory mechanisms, ensuring the emergence of virtuous behavior of actors regarding the use and management of data. These regulatory mechanisms rely firstly on the existence of *censorship* procedures against editors responsible for voluntarily downgrading the contents, and of *grants* of temporary extension of rights for trusted editors *[110], [111]*. These regulatory mechanisms formalize a hidden hierarchical structure underlying Wiki communities *[112]*. Another key requirement to promote quality of content emerging from free interactions is the *transparency* and *traceability* of the editors' actions, which help to develop confidence of actors and emancipatory effects of the system *[97]*. Cardon and Levrel *[113]* use the terms of *participatory vigilance* to describe Wikipedia's governance and its procedural system of self-regulation. It is a commonly accepted fact that community-based paradigm has demonstrated its undeniable ability to provoke changes in actors' behavior regarding data governance issues, including *self-regulation* and *emancipation*.

---

[5] *editors* are (1) sincerely engaged in a collaborative research of truth (2) through a formalized framework (3) excluding usage of the force (4) and offering the ideal conditions for the dialog between actors (5) staying open continuously and for a large period of time *[97]*

[6] Lee et al. *[105]* propose that "a virtual community is a technology-supported cyberspace, centred upon the communication and interaction of participants to generate member-driven information and knowledge, resulting in the building of interpersonal relationships".

Lastly, a key factor for the development of these schemes is their ability to get their actors use a universal common digital environment, adapted to the management of interactions within communities and offering associated services of interest.

Despite their efficiency, these systems remain, in our opinion, irrelevant to guarantee the conditions for deploying efficient data governance in extended organizations, due to numerous reasons:
- The data quality produced by the crowd in the community paradigm has been strongly criticized *[93]*. Task conflicts within the group generate both positive and negative effects on the produced content *[96], [112]*. Lebraty and Lobre *[114]* even argue that it is a variable-sum game, sometimes negative, sometimes positive. This makes it difficult to consider that the *norms* relevant to transverse reference data could emerge from free interactions within a community.
- The roles assigned to members within a community are self-regulated by the community, including content-oriented or administration-oriented roles *[96], [115]*. However, the importance of decisional issues in extended organizations requires data governance scheme to strengthen accountability and empowerment of actors inheriting structuring roles on a large part of strategic data. The use of self-regulated control channels and the lack of transparency and responsibility of the authors *[116]* are then a major obstacle to develop data governance framework based on community paradigm in extended organizations.
- The discrepancy between priority levels assigned to a same collection of data, respectively by top-level managers of the organization and by virtual community members concerned by these data, has critical consequences on the control channels efficiency.

As a result, the Community Paradigm, despite its undeniable success, cannot by itself provide a complete answer to the problem of finding a data governance scheme adapted to extended organizations.

### 2.3 Guidelines for a new data governance paradigm : *Panopticon*

MDM paradigm has been developed along the lines of preceding technical developments and existing IS architectures (ERP, BMPS, ETL, DataWarehouse). The *Panopticon paradigm* requires the development of new tools and architectures to articulate *regulatory* and *disciplinary* mechanisms to achieve effective data governance. This articulation is made concrete through a subtle action on representations relied on by the calculation of interests by the stakeholders, shared through their monitoring interfaces, and a control of the accountability and empowerment of the actors. This paradigm inherits main contributions from the *community* paradigm, but aims to compensate for its shortcomings. We propose the IS architecture of the new paradigm to be based on the existence of a specific IS element, called *Panopticon IS brick,* acting as a *hub* between existing elements of the organization's IS and personal digital environments of the individual. *Panopticon* is intended to formalize the architecture of visibility and power within the organization.

Deployment of the *Panopticon* data governance paradigm led to a radical transformation of business practices and address basic questions that should be analysed in following: the explicit construction of *area of visibility* and *fields of action* for individuals, the management of transversality between business areas and across boundaries of organization.

### 2.3.1 Power and visibility - Monitoring interfaces

The *Panopticon paradigm* must confer a central role to *monitoring interfaces* opened to users through their personal digital environment. The *user interface* has to offer to each individual a complete overview on the services he can access to, but also on the rights and responsibilities accorded to him on a selected collection of data. Different individuals are sharing the access to services and inherit potentially conflicting responsibilities on data, the *field of action* offered by the *user interface* becomes intrinsically an *area of visibility* for the other actors to monitor each action realized by the user. The way this *user interface* links the set of services in access to the user, on one hand, to the selected collection of data on which these services are based, on the other hand, impacts strongly the calculation of its *interests* to exercise its power on data belonging to its of *field of action.* Hence, the adaptation of this *user interface* is the way the *government* can model the *representations* of the user, to promote the *regulatory* and *disciplinary* mechanisms relied on by its *liberal governmentality*. Thus, the fundamentals of *Panopticon paradigm* are the following:

- Individuals can contribute within their own customized digital environment to a set of control processes on data belonging to their *field of action*. The data are presented in their current state of reliability, facing the user with the *interpellation* of the system to *recognize* its *digital dual* world and then constitute himself as a *subject* by using its power to tell their truth on these data. Unlike MDM solutions working downstream of SI elements, like a *Extract-Transform-Load (ETL)* system acts towards a Data Warehouse, *Panopticon IS brick* maintains its reference databases through *real-time processes*.
- Complete *transparency* and *traceability* are ensured on the set of *required interventions* made from individual actions or external sources (proposition to change the value of a given reference datum, reasoned opinion emitted to conclude within a given control channel, arbitration control between divergent control channels). Each actor involved in a control channel is then placed in an *area of visibility* for an invisible community of actors, concerned by the same data, in order to promote *self-discipline* and integration of norms. However, *anonymity* can be ensured on *free contributions* devoted to the *warning* about erroneous data and *critical/ranking processes*, in order to promote *emancipation* of the individual with respect to the issue of managing data.
- This approach is *user-centric*, in the sense that the collection of reference data, covered by the data governance scheme, is chosen according to the set of data used by the set of digital services offered to the users. User-interface is constantly adapted to the currently used services in order to optimally leverage *personal interest* of the users to get them to participate to control processes. This interest relies on its need to access services based on up-to-date and personalized data, to cooperate with other members of his networks, to develop competitive strategies to access shared resources, or to exercise his responsibilities.
- Numerous *control channels* exist for any given datum, a control channel is indeed associated to any community concerned by the different usages of this datum. Each control channel is formalized by the allocation of *structuring roles* and *prioritized rights* about this datum to any individuals within this community: rights to read, rights to freely warn for an erroneous data, rights and responsibility to propose a modification of a datum, rights and responsibility to evaluate/control the propositions to change a datum made by other individuals, right and responsibility to arbitrate between divergent controls. The set of control channels formalized by the system encompass the whole set of *ties, controls* or *transactions*, inherited from *networks* and coalitions existing within the organization, as well as conflictual and competitive relationships, although these relations are generically transverse to hierarchical relationships of the organization.
- Unlike in MDM scheme where the control channels are initial parameters for the system, the *Panopticon paradigm* allows the communities to *self-organize* the control channels. This bias is imposed by the objective fixed by the system to take into account the complex dynamics of these networks. Modifications made by an individual, on the *hierarchical data* belonging to its *field of action*, contribute to change this *field of action*, as well as the *area of visibility* within which he is located, but also to modify or constrain those of the other individuals. In order to conciliate the multiplication of self-organized control channels and the efficiency of the whole control process, we have to impose basic requirements: unlike in community-based data governance schemes a unique control channel associated to hierarchical channel inherits the arbitration power on the final decision and responsibility to change the golden record, the whole set of control channels concerned by the same collection of data are ranking/censoring/granting each other according to the rights they have to act on hierarchical data corresponding to the details of the other control channels.

In order to illustrate preceding requirements, we propose a simple and explicit approach to model the *field of action* and *area of visibility* of any actor in the system. This model is quite independent on the type of the considered organization:
- The organization is made up a set of separate *entities*, these entities are of various kinds: individuals, resources, structures, products, external data sources or authorities... Organizational relationships are described by *connections* between these entities, the whole set of connections form a graph which dynamics formalizes elementary business processes within the organization. *Specific attributes* of the entities (*Master Data*), *connection attributes* associated to the connections between entities (*Hierarchical Data*), and the history of the modifications proposed by users for the latter (*Transactional Data*) are basic elements registered in the Reference Data repository maintained by the *Panopticon IS brick*.
- The system associates to any entity of the organization a community on its own account, it consists in the set of individuals connected with this entity: the *community of the entity*. The attributes attached to the connexions between individuals and other entities characterize the basic organizational links involving the individual within the organization[7]. The individual within the *community of an*

---

[7] the individual is connected to a resource if he makes use of it or contribute to its management, he is connected to a structure if he is member of it or even assumes a management role within it, he is connected to a project/product if he contributes to its realization or owns benefit from it, he is connected to an authority or an external data source if their jurisdiction encompass an arbitration power on their personal attributes,...

*entity* inherit prioritized rights on a selected *collection of data associated to this community of the entity*. The *field of action* of any individual is formed by the union of the set of *collections of data associated to the communities of the entities* he is connected to.
- The *collection of data associated to the community of an entity* is formed by: the *specific attributes* of the entity, *connection attributes* of the entity to its parent-entities on the graph fixing external constraints weighting on the management of the entity, the *connection attributes* of the entity to its child-entities on the graph fixing the internal nested mechanisms of control on its subunits, and finally the *connection attributes* relative to the connection of its child-entities to their other parent-entities on the graph when these data are relevant for assessing the constraints affecting the management of these child-entities by the community of the entity. *Connection attributes* between two entities of the organization are in fact integrated in the collection of data of many entities, this fact shows the way the system multiplies the control channels on these data. The specific attributes of the entities are subject to the control of the community of the entity but also to the control of some authorities or external data sources in order for the attributes to be subject of numerous control channels.
- The *rights*, granted to individual members of the *community of an entity*, on its own collection of data, are deduced from the roles accorded to these individuals within the community, but also from indices characterizing the peculiar position of this community within the control processes on these shared data. These indices are intended to: singling out the community owning the arbitration power within the set of control channels deployed by the communities concerned by these data, specify the hierarchical levels to be mobilized in the various concerned communities to carry out their own control process on these data, the rules governing the *grants* and *censorships* affecting the common rights accorded to members of the various communities concerned by these data.

As a result, the processes governing the changes of reference data relies on:
- the changes performed on data by individuals owning appropriate *rights* through their *monitoring interfaces* ;
- the propagation rules inherited by transitivity along the links of the graph of connections due to functional rules within the organization, or the adhoc constraints inherited from business rules.

### 2.3.2 Panopticon Paradigm and complexity of extended organizations

The *Panopticon paradigm* must provide solutions to the challenges raised by the various sources of complexity of extended organisms. Let us briefly discuss these issues.

The various digital services of the organization underly formalized business processes, covering the definition and characterization of a set of objects, a normative description of their connections, as well as specific business rules governing the evolution of these, through the interventions of users, managers, external data sources and authorities. The same entity is generically associated with objects involved in a wide range of business processes. Specific and connection attributes associated to entities are *shared* between various business areas and partner organizations. The bias adopted by *Panopticon paradigm* to respect the existing business processes and the existing urbanized IS makes of cross-reliability mechanisms and of the deployment of global standards the main issues in the design and integration process of the *Panopticon IS brick*. The modeling of business process and of basic entities must tend to universality with respect to the diversity of processes covered by the existing IS bricks, and integrate a significant part of the business rules.

The *Panopticon paradigm* must also provide appropriate answers to the issue of dealing with boundaries of the extended organization.

While the MDM paradigm is not well adapted to the integration of external data sources, they should be extensively used by Panopticon scheme. They must be considered as well as the *control channels* emerged from communities to anticipate improvements and remedy to the control processes, which do not meet the appropriate data quality threshold. The entities characterizations (respectively of the connection between individuals and other entities) must provide measurement of structural or cyclical ability of the *community of the entity* (respectively of individuals) to ensure control operations that are expected from it. These controls can be handled personally by the designed actors or through rights delegation towards other individuals or external data sources. Control channels and data sources are subject to a *ranking* process by comparison with the results of other channels on the same data.

The answer given by the *MDM/ERP paradigm* to the issue of fostering data exchange protocols between the organization and its partners is to impose a single integrative framework for business processes. By contrast, *MDM paradigm* neglects the existence of *internal boundaries* emerging within organizations from resistance strategies deployed by some of its sub-units. To deal with these two types of boundary problems, the strategy adopted by *Panopticon paradigm* should be to promote a "functional decentralization" of a significant part of the control processes through the development of a *distributed IS* architecture based on numerous instances of the *Panopticon IS brick*. Reference databases relied upon by these different instances of *Panopticon IS brick* have to be synchronized through a protocol formalized by the contractual link between the organisation and its subunits or partners. This strategy promotes the dissemination of norms underlying the reference databases of the *Panopticon IS brick*, at the cost of losing visibility on a part of control processes carried out within the subunits.

Together these points deserve a precise analysis of technical issues which exceeds the scope of this article and will be the object of a separate work.

## Conclusion

After having clarified the constraints on data governance schemes within extended organizations, it became apparent that the current paradigms underlying the Master Data Management solutions, or adopted by digital networks communities, do not meet them. An analysis of the regulatory and disciplinary controls within these extended organizations has led us to propose a new paradigm to meet the constraints weighting on the deployment of such a scheme, it requires technological developments that should be the object of a specific research.